\documentclass[prb,twocolumn,draft,amsmath,showpacs]{revtex4}
\usepackage{graphicx}
\usepackage{natbib}
\usepackage{color}
\input epsf

\begin{document}


\title{Anomalous frequency and intensity scaling of collective and local modes
in a coupled spin tetrahedron system}

\author{Kwang-Yong Choi}
\affiliation{Department of Physics, Chung-Ang University, 221 Huksuk-Dong, Dongjak-Gu,
Seoul 156-756, Republic of Korea}
\author{Hiroyuki Nojiri}
 \affiliation{ Institute for Materials
Research, Tohoku University, Katahira 2-1-1, Sendai 980-8577, Japan}
\author{Naresh S. Dalal}
\affiliation{ Department of Chemistry and Biochemistry, Florida State University
and National High Magnetic Field Laboratory, Tallahassee, Florida 32306-4390, USA}
\author{Helmuth Berger}
\affiliation{ Institute de Physique de la Matiere Complexe,
 EPFL, CH-1015 Lausanne, Switzerland}
\author{Wolfram Brenig}
 \affiliation{ Institute for Theoretical Physics,
 TU Braunschweig, D-38106 Braunschweig, Germany}
\author{Peter Lemmens}
 \affiliation{ Institute for Condensed Matter Physics,
 TU Braunschweig, D-38106 Braunschweig, Germany}

\date{\today}
\begin{abstract}
We report on the magnetic excitation spectrum of the coupled spin tetrahedral system
Cu$_{2}$Te$_{2}$O$_{5}$Cl$_{2}$ using Raman scattering on single crystals. The
transition to an ordered state at T$_{N}^{Cl}$=18.2 K evidenced from thermodynamic data leads to the evolution of distinct low-energy magnetic excitations superimposed by a
broad maximum. These modes are ascribed to magnons with different degree of localization and a two-magnon continuum. Two of the modes develop a substantial energy shift with decreasing temperature similar to the order parameter of other Neel ordered systems. The other two modes show only a negligible temperature dependence and dissolve above the ordering temperature in a continuum of excitations at finite energies. These
observations point to a delicate interplay of magnetic inter- and intra-tetrahedra
degrees of freedom and an importance of singlet fluctuations in describing a spin dynamics.
\end{abstract}



\maketitle


\section{Introduction}

Frustration and competition of magnetic interactions is one of the central
concepts in condensed matter physics.~\cite{Diep} It is related to unusual
ground states and (quantum)-criticality as well as exotic low-lying
excitations. The latter may reach a fascinating complexity due to the
dichotomy of local singlet versus collective magnon states. A prominent
example is found in the 3D pyrochlore lattice antiferromagnets (AF),
consisting of corner-sharing tetrahedra. Such systems have a macroscopic
number of classical ground states. Weak residual interactions of lattice
and orbital origin lift these degeneracies in the limit to low
temperatures.~\cite{Tchernyshyov} Furthermore, order-by-disorder effects may be observed.


A weakly coupled counterpart of the pyrochlore system with S=1/2 is realized in the
oxohalide Cu$_{2}$Te$_{2}$O$_{5}$X$_{2}$ (X=Br,Cl) and Cu$_{4}$Te$_{5}$O$_{12}$Cl$_{14}$
compounds.~\cite{Johnsson,Takagi} Four Cu$^{2+}$ clusters form a distorted tetrahedron,
which aligns in chains along the $c$ axis. The tetrahedra are separated by lone-pair
ions within the $ab$ plane that allow an easy modification of the important in-plane
exchange paths~\cite{Valenti} along oxygen and halogenoid ions using substitutions,
chemical modifications~\cite{Takagi} and pressure.~\cite{Kreitlow,Wang} Unlike the
pyrochlore system, each tetrahedron is isolated while still being coupled by
inter-tetrahedral couplings. From spin topology point of view the arrangement of
Cu$^{2+}$ realizes all prerequisites for quantum criticality.

Cu$_{2}$Te$_{2}$O$_{5}$X$_{2}$ shows an incommensurate magnetic ordering at
T$_{N}^{Br}$=11.4 K in X=Br and T$_{N}^{Cl}$=18.2 K in X=Cl. \cite{Lemmens01,Zaharko04}
The observed ordered magnetic moment 0.395(5) $\mu_B$ of Cu$^{2+}$ is strongly reduced
for X=Br, compared to the classical value of $\sim 1\mu_B$. \cite{Zaharko06} In
contrast, for X=Cl the moment of 0.88(1) $\mu_B$ is closer to the saturated one. Since
the unit cell volume decreases by $7\%$ as Br is substituted by Cl, the ratio between
intra-tetrahedra and inter-tetrahedra couplings seems to be a crucial factor for
understanding the respective magnetic behavior. Indeed, the {\it ab initio}
calculation~\cite{Valenti} shows that exchange paths vary with composition because the
Cl 3p orbital at the Fermi level is more strongly distorted towards the Cu 3d orbital
than the Br 4p orbital. Moreover, for X=Cl in-plane inter-tetrahedral diagonal
interaction is estimated to be nearly as strong as the intra-tetrahedral interaction.
With applied pressure T$_{N}^{Br}$=11.4 K is systematically reduced, implying a decrease
of the magnetic inter-tetrahedra coupling strengths and an enhanced degree of
frustration.~\cite{Kreitlow} The effect of symmetry can be studied comparing
Cu$_{2}$Te$_{2}$O$_{5}$Cl$_{2}$ with Cu$_{4}$Te$_{5}$O$_{12}$Cl$_{14}$ as in the latter
system the spin tetrahedra have a larger separation within the ab plane including an
inversion center.~\cite{Johnsson,Takagi} As a result it shows a more mean-field like
character of the magnetic properties.

In spite of the similar magnetic ordering structure of
Cu$_{2}$Te$_{2}$O$_{5}$X$_{2}$ with X=Br,Cl, the detailed magnetic
properties differ from each other. First, the effect of an external field
and pressure on T$_{N}$ is opposite.~\cite{Lemmens01,Crowe06} For X=Br,
$T_{N}$ decreases with increasing external field or pressure while for X=Cl
$T_{N}$ increases with increasing field or pressure. Second, thermal
conductivity differs from each other. The bromide shows a round maximum at
low temperature while the chloride displays a levelling-off followed by a
steep increase for temperature below 15 K.~\cite{prester,Sologubenko}
Third, inelastic neutron scattering (INS) measurements uncovered that the
two compounds show a marked difference in the temperature dependence of
magnetic excitations.~\cite{Crowe05} For X=Br, upon heating the intensity
of magnetic excitations decreases monotonically while undergoing no change
in lineshape. For X=Cl, however, the magnetic continuum shifts to lower
energy and then evolves to a quasielastic diffusive response above
T$_{N}^{Cl}$. This suggests that the nature of spin dynamics of both
compounds is unlike.

In previous Raman investigations of Cu$_{2}$Te$_{2}$O$_{5}$Br$_{2}$
\cite{Lemmens01,Lemmens02,Lemmens03} various magnetic excitations have been observed in the spin singlet channel that provided evidence for the presence of a longitudinal
magnon in this system. The latter is an important prerequisite for a proximity to quantum criticality.~\cite{Gros03,Jensen03} In contrast, the magnetic excitations of the X=Cl system have not been fully addressed due to the lack of sizable single crystals. To enhance our understanding of this weakly interacting tetrahedron system and to differentiate the spin dynamics of X=Br and Cl a thorough Raman spectroscopy
investigation of X=Cl is indispensable.

In this paper, we report dc magnetic susceptibility, high-field magnetization, and Raman
scattering measurements of large single crystals of Cu$_{2}$Te$_{2}$O$_{5}$Cl$_{2}$. The
anisotropic magnetization suggests that the ground state is given by a long-range
ordered state. However, we observe an intriguing richness of the magnetic Raman spectrum
that consists of four sharp peaks as well as of a weaker, broad continuum. The former
are interpreted in terms of magnon excitation. The latter is due to two-magnon
scattering, whose temperature dependence is indicative of a minor contribution from
localized fluctuations. The scaling of the modes points to a sizable
contribution of singlet fluctuations to a spin dynamics.


\section{Experimental setup}

Single crystals of Cu$_{2}$Te$_{2}$O$_{5}$Cl$_{2}$ were prepared by the halogen vapor
transport technique, using TeCl$_{4}$ and Cl$_{2}$ as transport agents. Magnetic
susceptibility was measured by a SQUID magnetometer (MPMS, Quantum Design). High-field
magnetization measurements were carried out by means of a standard inductive method. A
fast sweeping pulsed field was generated by a capacitor bank of 90 kJ.~\cite{Nojiri} The sample is directly immersed in liquid $^{3}$He to maintain a temperature of 0.4 K. Raman scattering experiments were performed using the excitation line $\lambda= 514.5$~nm of an Ar$^{+}$ laser in a quasi-backscattering geometry.~\cite{Lemmens05} A comparably small laser power of
0.1~mW was focused to a 0.1~mm diameter spot on the surface of the single crystal in
contact gas. The scattered spectra were collected by a DILOR-XY triple spectrometer and a nitrogen cooled charge-coupled device detector.

\section{Experimental results}
\subsection{Magnetic susceptibility and magnetization}
Figure~1 displays the temperature dependence of the magnetic susceptibility $\chi(T)$ in
a field $H=0.1$ T for H $\mid\mid$ c and H $\perp$ c axis, respectively. Our results
confirm earlier data. \cite{Johnsson,Lemmens01,Gros03,Jensen03,Jaglicic} With decreasing
temperature $\chi(T)$ shows a broad maximum around $T_{max}=24$ K. This is associated
with the onset of short-range magnetic ordering. Upon further cooling, $\chi(T)$
exhibits a kink around 18.2~K and then drops to a finite residual value as $T
\rightarrow 0$. The kink is identified with a transition to a long-rang ordered state at
T$_{N}^{Cl}$=18.2~K as evidenced by the $\lambda$-like anomaly of $d\chi/dT$ (see the
inset of Fig. 1).

\begin{figure}[th]
      \begin{center}
       \leavevmode
       \epsfxsize=8cm \epsfbox{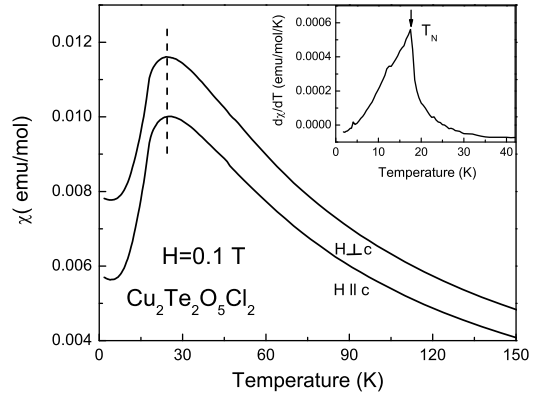}
        \caption{Temperature dependence of magnetic
susceptibility of Cu$_{2}$Te$_{2}$O$_{5}$Cl$_{2}$, $\chi(T)$,  in a an applied field of
$H=0.1$ T for H $\mid\mid$ c and H $\perp$ c axis, respectively. Inset: derivative of
magnetic susceptibility, $d\chi/dT$. The kink at T$_{N}^{Cl}$=18.2 K corresponds to
long-range magnetic ordering. } \label{Fig1}
\end{center}
\end{figure}

In Ref. \onlinecite{Johnsson} $\chi(T)$  was approximated in terms of an isolated
tetrahedral model with a spin-gapped state of $\Delta \approx J_1=J_2\sim 38.5$ K where
$J_1$ and $J_2$ are two intra-tetrahedral exchange interactions. However, it is
difficult to extract accurately the exchange interactions and spin gap from an analysis
of $\chi(T)$ since significant inter-tetrahedral couplings smear out the spin gap
features. \cite{Jensen03,Valenti,Whangbo,Brenig03} Actually, $\chi(T)$ becomes slightly
anisotropic for temperatures below $T_{max}$ due to the onset of long-range
correlations. In addition, $\chi(T)$ approaches a finite value as $T \rightarrow 0$
without falling to zero. This suggests that the spin gap is filled with dense
singlet-triplet mixed states leading to a finite magnetization.


\begin{figure}[th]
      \begin{center}
       \leavevmode
       \epsfxsize=8cm \epsfbox{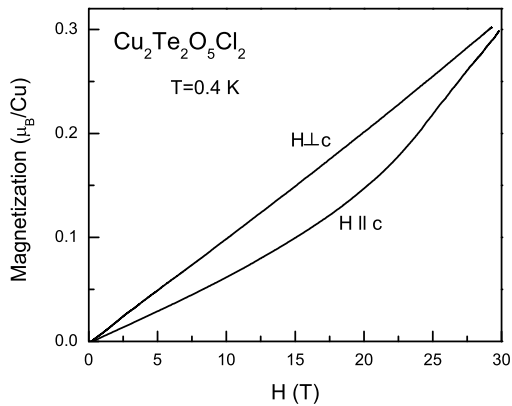}
        \caption{High-field magnetization of Cu$_{2}$Te$_{2}$O$_{5}$Cl$_{2}$
        for H $\mid\mid$ c and
H $\perp$ c axis at T=0.4 K, respectively,
        measured using a pulsed magnetic field.}
\label{Fig2}
\end{center}
\end{figure}

Shown in Fig. 2 is the high-field magnetization. We observe an anisotropic
magnetization behavior. For fields applied perpendicular to the $c$ axis,
the magnetization displays a linear field dependence, i.e. the
susceptibility is field independent. In contrast, for field applied along
the $c$ axis the magnetization is reduced with a concave curvature in the
studied field interval up to 30~T. We recall that a magnetization
plateau at half the saturation value has been predicted for a linear chain
of spin tetrahedra in a spin gapped ground state.~\cite{Totsuka02} The
absence of a half magnetization plateau together with the anisotropy
suggests that the ground state is governed by a classically ordered state
rather than by a spin singlet state. On a qualitative level, the
anisotropic magnetization behavior is compatible with helical magnetic
ordering. The linear field dependence for H $\bot$ c is associated with an
easy-plane type magnetization. The change of the magnetization slope for H
$\|$ might be related to a spin-flop transition with an incommensurate
wave vector.\\

\subsection{Raman scattering}
The low energy Raman spectra of Cu$_{2}$Te$_{2}$O$_{5}$Cl$_{2}$ are displayed in Fig. 3
for (cc), (aa), (ca), and (ab) polarizations at 3 K. We do not find any distinct
temperature dependence of the optical phonon modes in the frequency regime $80 - 700~
\mbox{cm}^{-1}$. Similar observations have been made for the other tetrahedra based
compounds.~\cite{Takagi,Lemmens01,Lemmens03,Gros03}

\begin{figure}[th]
      \begin{center}
       \leavevmode
       \epsfxsize=8cm \epsfbox{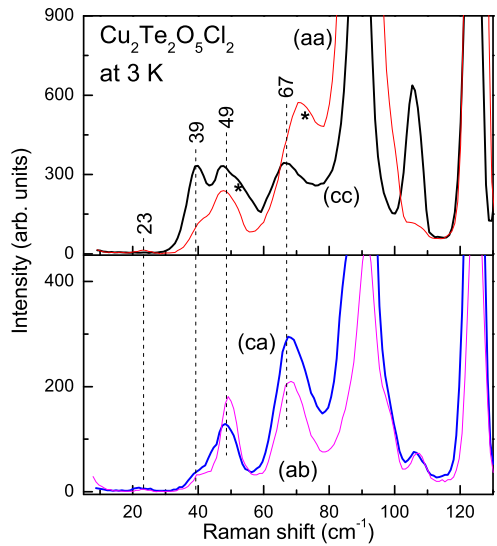}
        \caption{(Online color) Low-frequency Raman spectra of Cu$_{2}$Te$_{2}$O$_{5}$Cl$_{2}$ for
         (cc), (aa), (ca), and (ab) polarizations at T=3~K. The dashed lines
         denote the position of four magnetic signals. The numbers give their respective
         energy in the unit cm$\rm ^{-1}$. The asterisks denote low-frequency
         phonon modes that superimpose the magnetic signals.}
\label{Fig3}
\end{center}
\end{figure}

Hereafter, we will focus on the magnetic excitations which differ from the phonons by
their characteristic energy scale and the variation of both intensity and energy with
temperature. These excitations are composed of four peaks at 23 (P1), 39 (P2), 49 (P3),
and 67 cm$\rm ^{-1}$ (P4) as well as of a weak broad continuum (2M) extending from 30 to
120 cm$\rm ^{-1}$. With increasing temperatures an additional quasielastic signal (QC)
is observed. The modes are observed for all polarizations, indicating that spin
tetrahedra are networked in all three dimensions. However, also antisymmetric
Dzyaloshinsky-Moriya interactions may release the Raman scattering selection
rules.~\cite{Jensen03,Kotov05}


\begin{figure}[th]
   \begin{center}
       \leavevmode
       \epsfxsize=8cm \epsfbox{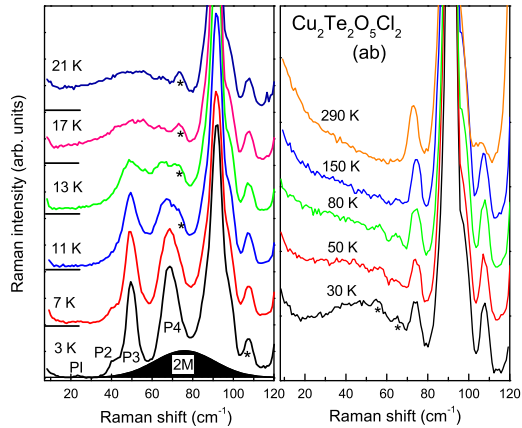}
        \caption{(Online color) Temperature dependence of low-frequency Raman spectra
        in (ab) polarization. The asterisks denote low-frequency phonon
         modes. Pi (i=1-4) and 2M correspond to the four sharp magnetic signals
         and the two-magnon continuum, respectively. }
\label{Fig4}
\end{center}
\end{figure}

\begin{figure}[th]
   \begin{center}
       \leavevmode
       \epsfxsize=8cm \epsfbox{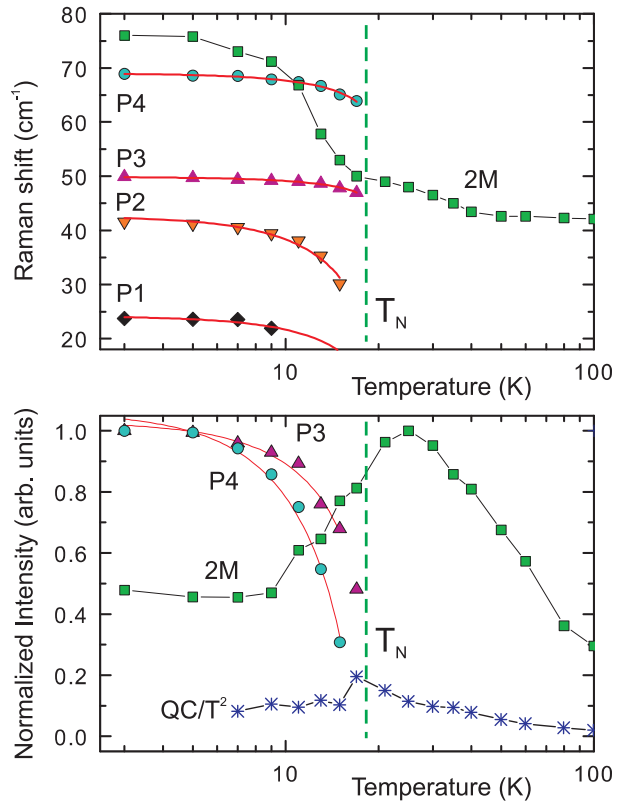}
        \caption{(Online color)
        (Upper panel) Peak frequency of the magnetic
        signals, Pi (i=1-4) and the two-magnon
        continuum, 2M for Cu$_{2}$Te$_{2}$O$_{5}$Cl$_{2}$ on a logarithmic temperature scale.
        (Lower panel) Scattering intensity
        of the magnetic signals. The intensity of quasielastic scattering,
        QC, is normalized by T$^2$ to be compared with the magnetic specific heat.~\cite{Lemmens01}
        Lines are guides to the eye. }
\label{Fig5}
\end{center}
\end{figure}

\begin{figure}[th]
   \begin{center}
       \leavevmode
       \epsfxsize=8cm \epsfbox{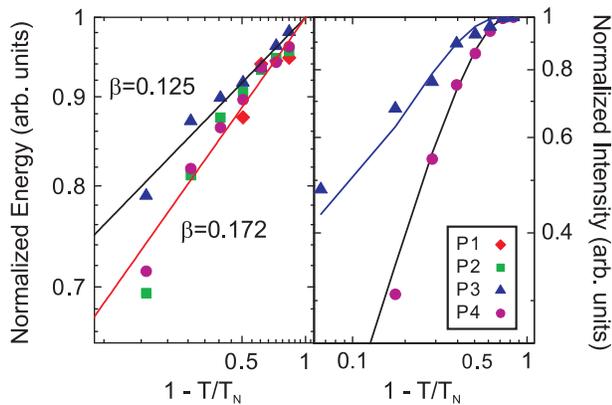}
        \caption{(Online color) Log-Log plots of the normalized energy and frequency of
magnetic scattering in Cu$_{2}$Te$_{2}$O$_{5}$Cl$_{2}$ as a function of the reduced
temperature, t=1-T/T$_N$. The results of the individual fits are given in Table I. (Left
panel) Normalized energy E-E$_0$/E$_{max}$ as a function of temperature for the modes
P1-P4. (Right panel) Normalized intensity, I/I$_{max}$, as a function of t for the modes
P3 and P4 compared to a Boltzmann related scattering factor (full lines) taking the
temperature dependence of the populated energy levels into account. } \label{Fig6}
\end{center}
\end{figure}

In Fig. \ref{Fig4} the detailed temperature dependence of Raman scattering data in
($ab$) polarization of Cu$_{2}$Te$_{2}$O$_{5}$Cl$_{2}$ is displayed. For a quantitative
analysis we fit them to Gaussian profiles after subtracting peaks with phonon origin.
The resulting peak frequencies and intensities are depicted in Fig.~\ref{Fig5} on a
logarithmic temperature scale and in Fig.~\ref{Fig6} on a log-log plot as a function of
the reduced temperature, t=1-T/T$_N$.

The intensity as well as the energy of modes P1-P4 are renormalized to a different
extent with increasing temperatures. This evolution can be used to characterize the
modes in addition to their absolute frequencies. While the intensities of P1 and P2 are comparably small and drop too fast to allow a detailed analysis the modes P3 and P4 show a moderate decrease of intensity following a more rapid drop in the proximity of
$T_{N}$. In contrast, the two-magnon scattering intensity increases and even forms a
maximum at 25~K, i.e. very close to the maximum in the magnetic susceptibility. This
increase of the scattering continuum resembles observations in the strongly frustrated,
2D Shastry-Sutherland system SrCu$_2$(BO$_3$)$_2$. In the latter system it is due to the localization of triplet excitations on a strongly frustrated lattice with a temperature
independent spin gap.~\cite{lemmens00} It is noteworthy that in the latter system the
intensity drops to zero for small temperatures while in the spin tetrahedron case finite intensity remains.

Dividing the intensity of magnetic quasielastic scattering ($\Delta\omega\approx0$) by
T$^2$ leads to a measure of the fluctuations of the magnetic energy density. This
quantity is proportional to the specific heat of a quantum spin system~\cite{Lemmens} as also observed in SrCu$_2$(BO$_3$)$_2$.~\cite{lemmens00} In the lower panel of Fig.~\ref{Fig5} this renormalized intensity is plotted for Cu$_{2}$Te$_{2}$O$_{5}$Cl$_{2}$
with a sharp maximum at $T_N$ which is similar to the earlier reported specific heat
data.~\cite{Lemmens01}

With respect to energy the peak frequencies of the modes P3 and P4 behave different from P1 and P2. The renormalization for T$<$$T_N$ is less pronounced and more step-like at T=$T_N$ resembling the effect of a first order like phase transition on the order
parameter. In contrast to P1-P4 the 2M energy does not soften completely from its low
temperature maximum of approximately 76~cm$\rm ^{-1}$. For temperatures above
$T_{N}^{Cl}$ there is still a finite energy spectral weight at 41 cm$\rm ^{-1}$ (see the upper panel of Fig.~5).

Comparing these observations with the generic behavior of local magnetic exchange
scattering of an AF system~\cite{cottam} leads to the following conclusions: With
respect to energy the sudden drop in energy and disappearance of the modes P3 and P4 is anomalous as well as the nearly constant energy of 2M for temperatures T$>$T$_N$. With
respect to intensity especially the behavior of 2M is noteworthy. In conventional AF the intensity of exchange scattering survives several times the N\'{e}el temperature without showing an anomalous enhancement. All our experimental results are consistent with the persistence of finite energy, short-range correlations, i.e. a gapped energy spectrum in the temperature regime T$>$$ T_{N}$.

In Fig.~\ref{Fig6} we show a scaling analysis of the peak energies and intensities by
log-log plots as a function of t=1-T/T$_N$.~\footnote{In Raman scattering on a large
semi-transparent crystal of CuGeO$_3$ and otherwise similar experimental conditions we
have estimated a local heating of the samples of 0.5~K/mW around T=14~K and an error of the temperature sensor of $\pm$0.5~K. We have tested the effect of a moderate change of T$_N$ in the power law fits leading to an approximate error of 10-20\% in $\beta$.} The full lines in the left panel correspond to a critical behavior with the exponents
$\beta$=0.125 and 0.172 in (E-E$_0$)/E$_{max}$=t$^\beta$, with the temperature
independent energy offset or gap E$_0$. It is clear from the previous discussion that P3 and P4 can only be described taking a finite offset E$_0$ into account. The same would be valid for the 2M energy, however, we have omitted this signal from the analysis because of its broader line width. To evaluate a scaling of P1 and P2 this offset is not needed. Therefore the latter two modes show a seemingly larger renormalization in the available temperature range. The detailed results of the individual fits for T=0 are given in Table~\ref{Tab1}.

\begin{table}
\caption{Results of fitting the energy of the four magnetic peaks to a scaling function
of (E-E$_0$)/E$_{max}$=t$^\beta$}  \label{Tab1}
  \centering

\begin{tabular}{|c||c|c|c|c|}

  \hline
  Mode & E$_0$ + E$_{max}$ (cm$^{-1}$) & E$_0$ (cm$^{-1}$) & E$_{max}$ (cm$^{-1}$) &  $\beta$ \\
  \hline
  P1   & 25   & 0     & 25    & 0.147 \\
  P2   & 43.5 & 0     & 43.5  & 0.193 \\
  P3   & 50.1 & 39.1  & 11    & 0.125 \\
  P4   & 69.5 & 54.1  & 15.4  & 0.172 \\
  \hline

\end{tabular}
\end{table}


In the right panel of Fig.~6 we have analyzed the intensity of P3 and P4 omitting the
other two signals due to their smaller intensities and sharp drop with rising
temperatures. It is obvious that a description of the intensity using a power law is not
satisfying. Therefore we used a model that has been successfully applied to the
intensity of bound states in $\rm SrCu_2(BO_3)_2$.~\cite{lemmens00} In this approach the
modes are suppressed due to thermal fluctuations taking the occupation of the modes and
their temperature dependence into account. This leads to a fit with essentially one free
parameter, a scattering efficiency A, as a prefactor of the Boltzman term. The
normalized intensity of P3 and P4 is therefore proportional to:
$I_{B}(T) \propto (1- A \cdot \ e^{\frac{-\Delta(T)}{k_B T}}~)~,$
with $\Delta(T)$ given by the temperature dependent peak energies P3 or P4, with A=9 and
0.5, respectively. This indicates that the rigidity of the modes P3 and P4 is determined
by thermal fluctuations as the bound states of $\rm SrCu_2(BO_3)_2$ do. Such a behavior
is not characteristic of the magnetic modes related to an order parameter.

We relate the temperature dependence peak energies to the evolution of inter-tetrahedral
correlations with the onset of ordering at T$\leq$$T_{N}^{Cl}$. The transversal triplet
excitation observed in neutron scattering on the spin chain system CuGeO$_3$ below its
spin-Peierls transition shows a similar behavior.~\cite{nishi-cugeo}

In contrast, the corresponding dependence of P3 and P4 corresponds to a smaller critical
exponent $\beta$. Therefore, we do not attribute these modes to the order parameter but
to spin singlet fluctuations arising from isolated spin tetrahedra. The corresponding
energy scale is not so well established and band structure calculations point to a
similar order of magnitude of the inter-tetrahedral interactions.~\cite{Valenti} A rough
approximation could be given by a fit to the magnetic susceptibility with an isolated
tetrahedron model leading to J=38.4~K.~\cite{Johnsson,Takagi} For X=Br there exists also
a high energy mode with very weak temperature dependence. This triangular, broadened
mode might be understood as composed of several modes, i.e. an overlap of P3 and P4 of
Cu$_{2}$Te$_{2}$O$_{5}$Cl$_{2}$. Although the higher energy modes in the two systems
show several similarities there is one major difference in the behavior at higher
temperatures, T$>$T$_N$. While the broad triangular mode for X=Br survives several times
T$_N$, the sharp modes for X=Cl rapidly disappear and dissolve in the broader 2M signal.
The interpretation of this difference is not straightforward and could be due to a
weakly first order contribution to the phase transition in
Cu$_{2}$Te$_{2}$O$_{5}$Cl$_{2}$.


Anticipating the later detailed discussion, we summarize that our study unveils a
survival of zero-dimensional quantum fluctuations attributed to individual spin
tetrahedra even though at lower temperatures long range ordering takes place due to the coupling of the spin entities.

\subsection{A comparison of neutron and Raman scattering}
In the following we will compare Raman scattering [Fig.~4 of Ref.
\onlinecite{Lemmens01}] with neutron scattering [Fig.~7(a) of Ref.
\onlinecite{Crowe05}] addressing
first the results for X=Br. Two components of the magnetic excitation spectrum have been observed in neutron scattering; (i) a flat, constant energy component and (ii) a
dispersive excitation. In a coupled tetrahedra system localized, dispersionless
excitations are expected to occur due to intra-tetrahedral interactions and dispersive
excitations due to inter-tetrahedral coupling. In this light, the former is related to a
spin gap feature while the latter to an incommensurate magnetic ordering. The
simultaneous observation of two components points to the coexistence of long range order
with a spin-gapped ground state. As discussed before, in Raman scattering a strongly
temperature dependent and a weakly temperature dependent feature exist. The small shift
of the latter signal is due to the damping of a Goldstone mode. Since the softened
spectral weight is small, we conclude that the ground state and spin dynamics of X=Br is
governed by spin singlet fluctuations. This is supported by the strongly reduced
magnetic moment 0.395(5)~$\mu_B$ of Cu$^{2+}$.~\cite{Zaharko06} For temperatures above
T$_{N}^{Br}$ the intensity of the continuum is monotonically suppressed without any
change in lineshape.

The similarity of the spectral response in INS and Raman scattering is a striking
feature considering different mechanisms for the scattering processes. Raman
spectroscopy probes simultaneous two-spin flip processes leading to a two-magnon
continuum. Thus, the magnetic continuum is proportional to twice the magnon density of
states. In contrast, INS corresponds to a spin-spin correlation function in momentum
space. As the available INS experiments have been performed on polycrystalline samples
the close correspondence between the two spectroscopic results might be based on the
averaging of the INS intensity over momentum space. This intensity is roughly given by
the one-magnon density of state.

Next, we turn to the discussion of the chloride. INS shows a magnetic excitation
spectrum that again consists of two components, that is, a flat, dispersionless band at
6~meV (48~cm$\rm ^{-1}$) and a dispersive lower energy component. Its energy scale is
smaller, 3~meV (24~cm$\rm ^{-1}$) and has a gap of 2~meV
(16~cm$\rm^{-1}$).~\cite{streule06} INS on a polycrystalline sample shows a progressive
shift of spectral weight to lower energy with increasing temperature and its transfer to
a QC diffusive response in the paramagnetic state [compare Fig.~4 and Fig.~7(b) of
Ref.~\onlinecite{Crowe05}]. More recently it has been shown that the dispersive mode
partially softens and remains gapped while being further broadened.~\cite{streule06} The
softening of the magnetic excitation implies the damping of short-range magnetic
fluctuations by thermal fluctuations.~\cite{Choi04} Therefore, we conclude that the spin
dynamics of X=Cl is dominated by long-range ordering in contrast to the case with X=Br.
This is consistent with the larger ordered moment of 0.88(1)~$\mu_B$ in X=Cl.

\subsection{Analysis of the magnetic Raman scattering}

Here we will discuss several options for the potential origin of the four sharp peaks at 23 (P1), 39 (P2), 49 (P3), and 67 cm$\rm ^{-1}$ (P4). As shown in Fig.~5, the peaks
shift to lower frequency with increasing temperature and vanish below the magnetic
ordering temperature. Thus, they might originate from a transverse magnon excitation at $q=0$.

Another possible interpretation  is in terms of a longitudinal magnon. Such excitations
have been observed in Raman scattering experiments of the sister compound of X=Br. To be more specific we start by considering the eigenstates of the isolated tetrahedron at site $\mathbf{r}$ with Hamiltonian
$H\left(\mathbf{r}\right)=J_{1}\left[\left(\mathbf{S}_{\mathbf{r}1}+\mathbf
{S}_{\mathbf{r}2}\right)\cdot\left(\mathbf{S}_{\mathbf{r}3}+\mathbf{S}_{
\mathbf{r}4}\right)\right]+J_{2}\left(\mathbf{S}_{\mathbf{r}1}
\cdot\mathbf{S}_{\mathbf{r}2}+\mathbf{S}_{\mathbf{r}3}\cdot\mathbf{S}_{
\mathbf{r}4}\right)$. These consist of two singlets $s_{1,2}$, at least one of which is the ground state, three triplets $t_{1,2,3}^{\alpha}$, and one quintuplet $q^{\alpha}$.
At $J_{1}=J_{2}$ the two singlets form a degenerate ground state. While $J_{1}\approx
J_{2}$ applies to X=Cl, we assume $J_{2}<J_{1}$ for definitness and $s_{1}$ to be the
ground state. This is similar to X=Br. The actual structure of the inter-tetrahedral
coupling in the tellurates is an open issue, but to a first approximation may be modeled by a molecular field
$H_{MF}=\sum_{\mathbf{r}l}\mathbf{M}_{\mathbf{r}l}\cdot\mathbf{S}_{\mathbf{ r}l}$ with
some incommensurate order parameter $\mathbf{M}_{\mathbf{r}l}$ for
$T<T_N$.~\cite{Zaharko04,Zaharko06} As suggested by the analysis in Refs.
\onlinecite{Gros03} and \onlinecite{Jensen03}, $H_{MF}$ will mix the ground state
singlet with the triplet components along the local quantization axis of the molecular
field. \emph{A} \emph{priori}, such mixing does not have to be restricted to only one of the triplets, as in Ref.~\onlinecite{Gros03}, but may involve all three. Discarding the high-energy quintuplet this would imply four low-energy excited states directly
observable at zero momentum transfer by Raman scattering, i.e. one singlet $s_{2}$ at
energy $2J_{1}-2J_{2}$ and three longitudinal magnons which vanish above $T_{N}$. The
$3\times2$ transverse triplets contribute to the two-magnon scattering continuum only.
This sets in at higher energies and remains unaffected by the the transition at $T_N$
-as for the remaining one- and two-magnon contributions from the quintuplet.

From this one should conclude first, that one of the four modes P1-P4 corresponds to the
excited singlet. In comparison to the 23.2~cm$^{-1}$ mode for X=Br, P1 would be a likely
candidate for this, leaving P2-P4 for the longitudinal magnons. Second, the singlet mode
among P1-P4 should be affected only weakly by an external magnetic field, while the
remaining three should be field-dependent through their triplet admixture. This should
be investigated by future Raman studies in finite magnetic fields. Finally, for a second
order transition the longitudinal magnon energies should vanish as $T\rightarrow T_{N}$
for $T< T_{N}$. However, while some softening is observable in all of the modes in
Fig.~\ref{Fig5}, their behavior would be more indicative of a weakly first order
transition. Indeed and depending on the molecular field, first order transitions may
occur for coupled tetrahedra - as already noted in the comment under Ref.~${12}$ in
Ref.~\onlinecite{Gros03}.


\section{Conclusions}
To conclude, we have presented a magnetic susceptibility, high-field magnetization, and Raman scattering study of the coupled spin tetrahedral system
Cu$_{2}$Te$_{2}$O$_{5}$Cl$_{2}$. Several distinct magnetic excitations are observed as
one-magnon modes in addition to a two-magnon continuum. The exceptionally rich magnetic excitation spectrum evidences the significance of a localized spin singlet dynamics
arising from zero-dimensional spin tetrahedra topology even though the static ordered
moment has nearly a classical value.

\section*{Acknowledgements}
This work was supported by the German Science Foundation and the ESF program
\emph{Highly Frustrated Magnetism}. Work at the EPFL was supported by the Swiss NSF and
by the NCCR MaNEP. Work at FSU was supported  by NSF (DMR-0506946). We acknowledge
important discussions with R. Valent\'{\i}.

\end{document}